\documentclass[conference]{IEEEtran}
\usepackage{graphicx}
\usepackage{amsmath}
\usepackage{pdfsync}

\usepackage{xcolor}
\usepackage{graphicx}

\begin{document}

\title{Towards Adaptive Resilience in\\
	High Performance Computing}

\author{\IEEEauthorblockN{Siavash Ghiasvand}
\IEEEauthorblockA{Center for Information Services\\
and High Performance Computing\\
Technische Universit\"at Dresden, Germany\\
siavash.ghiasvand@tu-dresden.de}
\and
\IEEEauthorblockN{Florina M. Ciorba}
\IEEEauthorblockA{Department of Mathematics\\
and Computer Science\\
Universit\"at Basel, Switzerland\\
florina.ciorba@unibas.ch}
}

\maketitle

\IEEEpeerreviewmaketitle

With the current growth in computing capabilities of high performance computing (HPC) systems, Exascale\footnote{Capable of performing $10^8$ floating point operations per second.} HPC systems are expected to arrive by 2020~\cite{nagel3}.
As systems become larger and more complex, they also become more error prone~\cite{Schroeder2007}.
The failure rate of HPC systems rapidly increases, such that, failures become the norm rather than the exception.
Therefore, in such unreliable environment, to maintain HPC systems operational, they must be resilient to failures.
Different approaches in HPC systems have been introduced to prevent failures (e.g., redundancy) or at least to minimize their impacts (e.g., via checkpoint and restart).
In most cases, when these approaches are employed to increase the resilience of certain parts of a system, performance significantly degrades, and/or energy consumption rapidly increases.

In general, failures can be divided into two groups: avoidable and unavoidable.
Since there is no `eternal' hardware, in theory, failures can not be truly `avoided'.
However, one can significantly decrease the probability of their occurrence, or in some cases postpone them for a certain amount of time.
Avoidable failures are defined in this work as failures that can be hidden from a specific system layer.
In contrast, the unavoidable failures are failures that cannot be hidden from a specific system layer.
We analogize the failure sources in an HPC system to a tree, which has its roots in the lowest system layer and its leaves in the highest layers.
As we traverse the system from top to bottom, the diversity of failures decreases while their impacts increase.
During propagation across system layers, failures may retain their original characteristics or they may morph into other types of failures.
Therefore, each system layer requires its own protection to prevent the propagation of specific types of failures to the upper layers.

While protection layers are added between system layers to identify, address, and prevent failures from propagating upwardly, certain overheads are imposed on the system.
As long as the failure protection layers are in place, they impose overheads, regardless of the presence or absence of failures.
In certain cases, adding overheads might not be worthwhile to provide fault tolerance.

This work in progress, proposes an approach that employs a probabilistic failure predictor to estimate the situations in which failures will occur in future.
Based on the probability of failures, it can be decided whether the available failure protection layers need to be activated or not .
Via modeling general HPC systems and estimating their failure rate, it was observed that applying a logical topology, may reduce failure probability, and that the logical topology also constitutes a uniform topology for any HPC system, such that the proposed failure probability predictor can be applied.
There are three main goals to achieve using the proposed approach:
Improved resilience, progress in computation, and energy saving.

In this approach, the HPC system is considered in its entirety and resilience mechanisms (e.g., checkpointing, isolation, and migration) are activated on-demand.
Using this approach, the unavoidable increase in total system performance degradation and energy consumption is decreased compared to the typical checkpoint/restart and redundant resilience mechanisms. 
Our work aims to mitigate a large number of failures occurring at various layers in the system, to prevent their propagation, and to minimize their impact, all of this in an energy-saving manner. 
In the case of failures that are estimated to occur but cannot be mitigated using the proposed approach (e.g., no surrogate resource is available), the system administrators will be notified in view of performing further investigations and reactions.

A resilient HPC system is a system which can complete the users requests even if certain units\footnote{A unit can be any component of the system, from a single transistor to an entire rack of computers, depending on the assumed component granularity.} of the system encounter failures and are no longer functioning.
A failure is a complete outage of a given system unit.
Thus, a malfunctioning or misbehaving unit is not a failed unit and, therefore, it is beyond the scope of this work.
Failures and their \textit{impact propagation} have an \textit{effectiveness zone}.
After each failure one can expect new failures, as part of the \textit{impact effectiveness zone}.
A sequence of successive failures is called a \emph{failure chain}.
Failures always propagate horizontally within a single layer, as well as from bottom to top across the horizontal system layers.

The common architecture of today's computers is based on the \textit{von~Neumann} description, first introduced by John von Neumann in 1945~\cite{VonNeumann1945, 194088}.
In the \textit{von Neumann} model, the computer consists of three main components:
\emph{central processing unit} (CPU), \emph{memory}, and \emph{input/output devices} (I/O).

HPC systems are also clusters of computing nodes which are connected together via a network of switches.
Therefore, the entire HPC system can also be modeled via \textit{von~Neumann} basic components.
There are different physical topologies to connect computational nodes together and form an HPC system.
The fat-tree topology is one of the most popular topologies~\cite{Top500}.
The indirect bidirectional multistage topology of fat-tree, given its symmetric layered formation and lack of \textit{root bottleneck}, has a natural fault-tolerant property.
In specific HPC systems, the numbers of nodes, chassis, and racks are chosen based on certain application-, budget-, and space-related considerations to increase their performance and reduce the acquisition costs.
However this may decrease the innate resilience of the fat-tree topology.
As stated earlier, an HPC system can be modeled via \textit{von~Neumann} basic components.
To reduce the system complexity and to facilitate a generic model it is assumed that all components have equal importance and that the impact of failure of one component on other components is instantaneous, which in reality may not be always the case.
The three basic components of \textit{von~Neumann} description, are serially connected to each other.
However, a computer may also have several parallel units.
In such cases, we can assume all similar parallel components as one super-component.
Based on these assumptions, the failure rate of a computer can be estimated via Eq.~(\ref{eq:node-failure-probability-general}), in which $FCMP_{cmp}$ denotes the estimated failure probability of a single component ${cmp}$.
\begin{equation}
	FN_n=(1-\prod^{last}_{cmp=first}(1-FCMP_{cmp}))
	\label{eq:node-failure-probability-general}
\end{equation}
Each chassis of an HPC system consists of a set of parallel computers (hereafter, \emph{node}).
Each rack is a set of chassis, and the HPC system is a set of racks.
Based on this topology, chassis is considered to be "failed" when all of its internal nodes fail.
The same definition applies to the racks and the HPC system.
Thus, the proposed failure estimation model can be recursively expanded from a single node, to chassis, to racks, and to the entire HPC system.

Applying this model to the statistics obtained from ~\cite{Schroeder2007} led us to the following set of results:
(1)~The failure impact of nodes' internal components (e.g., CPU, and memory), in comparison with the failure impact of external components (e.g., network switches) on the whole system failure rate is negligible; 
(2)~Having less than 4 nodes in a chassis is not efficient in the fat-tree logical topology;
(3)~Having more than 4 nodes in a chassis has no significant impact on reducing failure probability; 
(4)~In the fat-tree logical topology, using more branches with fewer leaves on each branch is more efficient than using few branches with many leaves on each branch;
(5)~In the fat-tree logical topology only the two top most component layers have significant impact on the whole system failure probability.

Repeating the calculations on higher system layers, and changing the granularity, provided the following results: 
(1)~Redundancy in higher layers is more beneficial;
(2)~Having less than 4 chassis in a rack is not efficient for failure prevention;
(3)~Having more than 8 chassis in a rack has no significant impact on decreasing failure probability;
(4)~Having more than 2 racks improves the failure rate;
(5)~Having more than 8 racks has no significant impact on failure reduction;
(6)~Expectedly the shared resources (e.g. network switches) have the dominant impact on the resilience of HPC system.

Based on these results, the proposed approach:
(1)~Applies a 4-4-4 logical fat-tree topology to the HPC system;
(2)~Based on the \textit{von~Neumann} model of HPC system, it estimates the system failure probability; and
(3)~Upon detection or prediction of failures, it activates failure protection layers within the \textit{effectiveness zone} of that failure;
(4)~Via this approach the formation of the \textit{failure chains} in the HPC system is prevented or minimized.
Via these steps, this approach can provide \textit{on-demand resilience} for HPC systems, which paves the way towards \textit{adaptive resilience}.

In general, this work makes the following contributions.
(1)~Proposes an approach to reduce the cost of resilience in HPC systems.
(2)~Recommends a logical topology which takes advantage of the built-in resilience of tree-like topologies, and increases system resilience at minimum overhead.
(3)~Proposes a general model to estimate failure probability in HPC systems based on \textit{von Neumann} model.
(4)~To the best of our knowledge, this paper is the first attempt to use logical topology to decrease the cost of resiliency in HPC systems.
(5)~It is also, to the best of our knowledge, the first study which uses system-wide failure probability estimation as the decision factor to provide on-demand resilience by controlling failure protection mechanisms (e.g., reconfiguring the checkpoint/restart and redundancy mechanisms) on HPC systems.

For the future work, beside improving the failure prediction model based on failure correlations~\cite{Ghiasvand2016}, analyzing the propagation pattern of failures in different system layers, and quantifying failures' impact within their \textit{effectiveness zone} has been planned.

\bibliographystyle{./style/IEEEtranBST/IEEEtran}
\bibliography{literature}

\end{document}